\documentclass[10pt,a4paper,twocolumn]{article}

\usepackage[utf8]{inputenc}
\usepackage{authblk}
\usepackage{graphicx}
\usepackage[explicit]{titlesec}
\usepackage[labelfont=bf,labelsep=endash,font=footnotesize]{caption}
\usepackage{tabu}
\usepackage{xcolor}
\usepackage[hidelinks]{hyperref}
\usepackage[hang]{footmisc}
\usepackage[normalem]{ulem}
\usepackage[top=2.5cm, bottom=2.8cm, left=1.5cm, right=1.5cm]{geometry}
\usepackage{abstract}
\usepackage{mathtools}
\usepackage{balance}

\usepackage{multirow}

\makeatletter
\renewcommand\AB@affilsepx{, \protect\Affilfont}
\makeatother

\providecommand{\keywords}[1]{\textbf{Keywords}\ \ \textendash\ \   #1}

\titleformat{\section}{\large\bfseries}{\thesection.}{1em}{\MakeUppercase{#1}}
\titlespacing*{\section}{0pt}{12pt}{6pt}

\titleformat{\subsection}{\large}{\thesubsection}{1em}{#1}
\titlespacing*{\subsection}{0pt}{12pt}{6pt}

\titleformat{\subsubsection}{\large\itshape}{\thesubsubsection}{1em}{#1}
\titlespacing*{\subsubsection}{0pt}{12pt}{6pt}

\newcommand{\ITUurl}[1]{\textcolor{blue}{\urlstyle{same}\url{#1}}}

\newcommand{\ITUpar}{\vspace{8pt}\par}

\newcommand{\ITUfootnote}[1]{\footnote{#1}}

\renewenvironment{abstract}
               {\list{}{
               \setlength{\rightmargin}{0mm}
               \setlength{\leftmargin}{0mm}
               \vspace{-0.25in}
                \item[\textit{\textbf{\hspace{22pt}Abstract  }}  \textendash]\relax}}
               {\endlist}
\def\starttable{\vspace{6pt}\begin{table}[ht]\center}
\def\startfigure{\vspace{6pt}\begin{figure}[ht]\center}

\makeatletter
\def\tagform@#1{\maketag@@@{\ignorespaces#1\unskip\@@italiccorr}}
\makeatother

\usepackage{caption}
\usepackage{subcaption}

\title{\large{\textbf{\uppercase{Machine Learning for Performance Prediction of Channel Bonding in Next-Generation IEEE 802.11 WLANs}}}}

\author[1]{\normalsize{Francesc~Wilhelmi}}
\author[2]{\normalsize{David~Góez}}
\author[3]{\normalsize{Paola~Soto}}
\author[1]{\normalsize{Ramon~Vallés}}
\author[4]{\normalsize{Mohammad~Alfaifi}}  
\author[4]{\normalsize{Abdulrahman~Algunayah}}
\author[5]{\normalsize{Jorge~Martin-Pérez}}
\author[5]{\normalsize{Luigi~Girletti}}
\author[6]{\normalsize{Rajasekar~Mohan}}
\author[6]{\normalsize{K~Venkat~Ramnan}}
\author[1]{\normalsize{Boris~Bellalta}}

\affil[1]{\normalsize{Universitat~Pompeu~Fabra}}
\affil[2]{\normalsize{Universidad~de~Antioquia}}
\affil[3]{\normalsize{University~of~Antwerp}}
\affil[4]{\normalsize{Saudi~Telecom}}
\affil[5]{\normalsize{Universidad~Carlos~III~de~Madrid}}
\affil[6]{\normalsize{PES~University}}

\date{\vspace{-12pt}\endgraf\rule{\textwidth}{1pt}}

\begin{document}


\twocolumn[

\begin{@twocolumnfalse}
\maketitle

\begin{abstract}
\textit{With the advent of Artificial Intelligence (AI)-empowered communications, industry, academia, and standardization organizations are progressing on the definition of mechanisms and procedures to address the increasing complexity of future 5G and beyond communications. In this context, the International Telecommunication Union (ITU) organized the first AI for 5G Challenge to bring industry and academia together to introduce and solve representative problems related to the application of Machine Learning (ML) to networks. In this paper, we present the results gathered from Problem Statement~13 (PS-013), organized by Universitat Pompeu Fabra (UPF), which primary goal was predicting the performance of next-generation Wireless Local Area Networks (WLANs) applying Channel Bonding (CB) techniques. In particular, we overview the ML models proposed by participants (including Artificial Neural Networks, Graph Neural Networks, Random Forest regression, and gradient boosting) and analyze their performance on an open dataset generated using the IEEE 802.11ax-oriented Komondor network simulator. The accuracy achieved by the proposed methods demonstrates the suitability of ML for predicting the performance of WLANs. Moreover, we discuss the importance of abstracting WLAN interactions to achieve better results, and we argue that there is certainly room for improvement in throughput prediction through ML.}
\end{abstract}
\ITUpar

\ITUpar
\ITUpar

\end{@twocolumnfalse}
]

\section{Introduction} 
\label{sec:intro}

The utilization of Artificial Intelligence (AI) and Machine Learning (ML) techniques is gaining momentum to address the challenges posed by next-generation wireless communications. The fact is that networks are nowadays facing unprecedented levels of complexity due to novel use cases including features such as spatial multiplexing, multi-array antenna technologies, or millimeter wave (mmWave) communications. While these features allow providing the promised performance requirements in terms of data rate, latency, or energy efficiency, their implementation entails additional complexity (especially for crowded and highly dynamic deployments), thus making hand-crafted solutions unfeasible.\ITUpar

IEEE 802.11 Wireless Local Area Networks (WLANs) are one of the most popular access solutions in the unlicensed band, and they represent a prominent example of increasing complexity in wireless networks. The optimization of WLANs underlines particular challenges due to the decentralized nature of these types of networks, which mostly operate using Listen-Before Talk (LBT) transmission procedures. If to this we add that WLAN deployments are typically unplanned, dense, and highly dynamic, the complexity is even increased.\ITUpar

To address the optimization of next-generation WLANs, the usage of AI/ML emerges as a compelling solution by leveraging useful information obtained across data, which allows deriving models from experience. Nevertheless, the adoption of AI/ML in networks is still in its initial phase, and a lot of work needs to be done. In this regard, standardization organizations are undertaking significant efforts towards fully intelligent networks. An outstanding example can be found in the International Telecommunication Union (ITU)'s ML-aware architecture~\cite{bib1}, which lays the foundations of pervasive ML for networks.\ITUpar

Another aspect essential for the prosperity of AI/ML in communications is data availability and openness. In this context, the \emph{ITU AI/ML in 5G Challenge}~\cite{bib2} was set in motion to encourage industry, academia, and other stakeholders to collaborate and exchange data for solving relevant problems in the field. This initiative entailed a big step forward in bringing open source closer to standards.\ITUpar

As a contribution to the ITU challenge, and aligned with WLANs optimization, in this paper, we present the results obtained from problem statement \textit{``Improving the capacity of IEEE 802.11 Wireless Local Area Networks (WLANs) through Machine Learning"} (referred to as PS-013 in the context of the challenge), whereby participants were called to design ML models to predict the performance of next-generation Wi-Fi deployments. For that purpose, an open dataset with WLAN measurements obtained from a network simulator was made available. As later discussed, network simulators are gaining importance to enable future ML-aware communications by acting as ML sandboxes. In the context of the challenge, synthetic data was used for training ML models.\ITUpar

In summary, based on the proposed problem statement and the solutions provided by participants, we discuss the feasibility of predicting the throughput of complex WLAN deployments through ML. To the best of our knowledge, this is an under-researched subject with high potential. Accurate performance predictions may open the door to novel real-time self-adaptive mechanisms able to enhance the performance of wireless networks by leveraging spectrum resources dynamically. For instance, given a change in the network configuration, predictions about future performance values can be used as a heuristic to guide the choices made by algorithms that operate in decentralized self-configuring environments \cite{bib3}.\ITUpar

Table~\ref{tab:tab0} briefly summarizes all the models proposed by the participants of the challenge and the main motivation behind them. For instance, the model proposed by the \textit{ATARI} team aims to exploit the graph structure inherent in WLAN deployments. Alternatively, the solution provided by \textit{Ramón Vallés} is focused on abstracting different categories of features (e.g., signal quality, bandwidth usage) and generate predictions based on them.\ITUpar

\starttable
\caption{Summary of the ML models proposed by the participants of the challenge.}\label{tab:tab0} 
\begin{small}
	\resizebox{\columnwidth}{!}{%
	\begin{tabular}{|c|c|l|c|}
		\hline
		\textbf{Team} & \textbf{Proposed Model} & \multicolumn{1}{c|}{\textbf{Motivation}} & \textbf{Ref.} \\ \hline
		ATARI & \begin{tabular}[c]{@{}c@{}}Graph Neural \\ Network\end{tabular} & \begin{tabular}[c]{@{}l@{}}Exploit graph \\ representation \\ of WLANs\end{tabular} & \cite{bib13}  \\ \hline
		\begin{tabular}[c]{@{}c@{}}Ramon \\ Vallés\end{tabular} & \begin{tabular}[c]{@{}c@{}}Feed-forward \\ Neural Network\end{tabular} &  \begin{tabular}[c]{@{}l@{}}Abstract problem \\ characteristics by \\ categories\end{tabular} & \cite{ramon_github}  \\ \hline
		STC & Gradient Boosting & \begin{tabular}[c]{@{}l@{}}High performance, \\ flexibility, and ease \\ of deployment\end{tabular} & \cite{github_stc} \\ \hline
		\begin{tabular}[c]{@{}c@{}}UC3M \\ NETCOM\end{tabular} & \begin{tabular}[c]{@{}c@{}}Feed-forward \\ Neural Network\end{tabular} & \begin{tabular}[c]{@{}l@{}}Learn throughput \\ function exhaustively\end{tabular} &  \cite{github_netcom}\\ \hline
		\begin{tabular}[c]{@{}c@{}}NET \\ INTELS\end{tabular} & \begin{tabular}[c]{@{}c@{}}Random Forest \\ Regression\end{tabular} & \begin{tabular}[c]{@{}l@{}}Address problem's \\ non-linearity and \\ reduce dimensionality\end{tabular} & \cite{netintels_github} \\ \hline
	\end{tabular}
	}
\end{small}
\end{table}

The results presented in this paper showcase the feasibility of applying ML to predict the performance of next-generation WLANs. In particular, some of the proposed models have been shown to achieve high prediction accuracy in a set of test scenarios. Moreover, we have identified the main potential and pitfalls of the proposed models, thus opening the door to new contributions that improve the baseline results shown in this paper. The dataset is available online~\cite{dataset}, and we expect it can be used for benchmarking other ML methods in the future.\ITUpar

The remaining of this paper is structured as follows: first, Section~\ref{section:ps} presents the CB problem for next-generation WLANs and describes the dataset provided for throughput prediction in dense deployments. Then, Section~\ref{section:solutions} overviews the ML-based solutions proposed by the ITU challenge participants, which results are provided in Section~\ref{section:performance}. Finally, Section~\ref{section:conclusions} concludes the paper with final remarks.\ITUpar

\section{Channel bonding in next-generation WLANs}
\label{section:ps}

In this section, we first describe the CB problem in WLANs and underscore the need for ML models for performance prediction. Then, we introduce the dataset provided for the ITU challenge, which is open to any researcher interested in this topic.

\subsection{Channel Bonding in IEEE 802.11 WLANs}

Next-generation IEEE 802.11 WLANs are called to face the challenge of providing high performance under complex situations, e.g., to provide high throughput in massively crowded deployments where multiple devices coexist within the same area. To fulfill the strict requirements derived from novel use cases, features such as Multiple-Input Multiple-Output (MIMO), Spatial reuse (SR), or multi-Access Point (AP) coordination are being developed and incorporated into the newest amendments, namely IEEE 802.11ax (11ax) and IEEE 802.11be (11be)~\cite{bib4,bib5}.\ITUpar 

One of the features that are receiving more attention is Channel Bonding (CB) \cite{bib6, bib7}, whereby multiple frequency channels can be bonded with the aim of increasing the bandwidth of a given transmission, thus potentially improving the throughput. Since its introduction to the 802.11n amendment, where up to two basic channels of 20 MHz could be bonded to form a single one of 40 MHz, the specification on CB has evolved and currently allows for channel widths of 160 MHz (11ac/11ax). Moreover, CB is expected to support up to 320 MHz channels in the 11be amendment.\ITUpar

\startfigure
\includegraphics[width=.75\columnwidth]{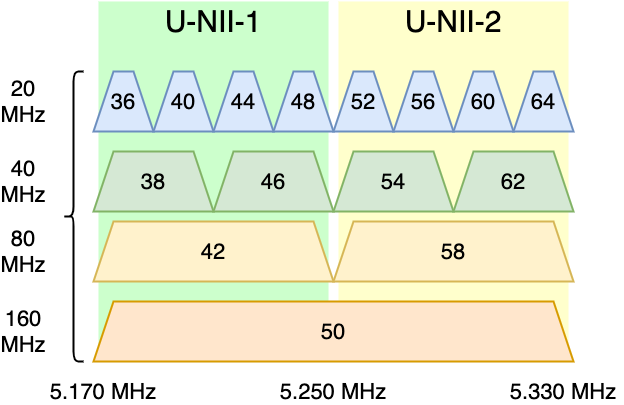}
\caption{U-NII-1 and U-NII-2 sub-bands of the 5 GHz Wi-Fi channels.}\label{fig:fig1} 
\end{figure}

Figure~\ref{fig:fig1} shows a snapshot of the 5~GHz band in Wi-Fi (particularly, U-NII-1 and U-NII-2 bands are shown), where basic 20 MHz channels can be bonded to form wider channels of up to 160 MHz. As it can be appreciated, the number of combinations for bonding channels is high, even for a small portion of the available spectrum\ITUfootnote{In the 5 GHz and 6 GHz bands, there are six and fourteen non-overlapping channels of 80 MHz, respectively.} and under the constraint that only contiguous channels can be bonded. Moreover, novel bonding techniques combine Orthogonal Frequency Multiple Access (OFDMA) with preamble puncturing~\cite{bib51} to use non-contiguous channels. With all this, given the number of Basic Service Sets (BSSs) and devices in crowded deployments (see Fig.~\ref{fig:fig2}), we can say that CB is a problem with a combinatorial action space.\ITUpar

\startfigure
\includegraphics[width=\columnwidth]{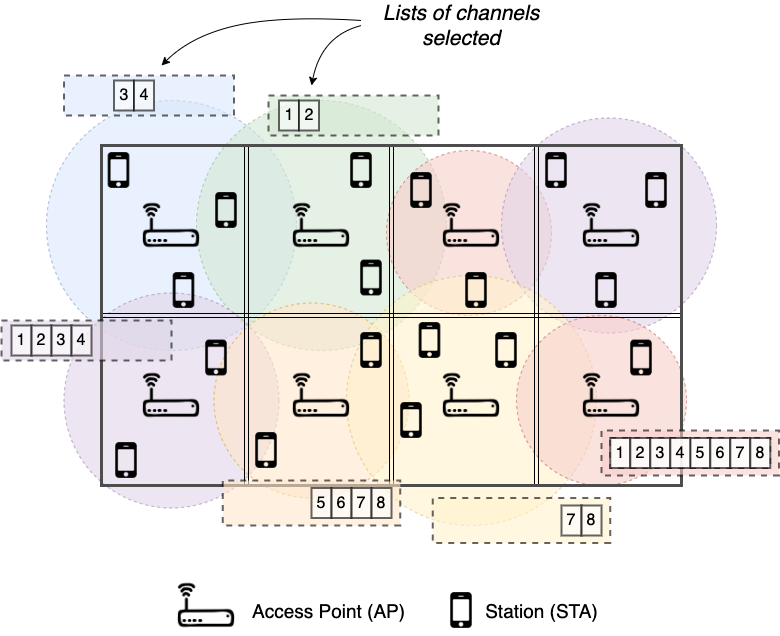}
\caption{Dense WLAN deployment with multiple CB configurations.}\label{fig:fig2} 
\end{figure}

\subsection{Policies for Dynamic Channel Bonding}

To harness the available spectrum within the CB operation, Dynamic CB (DCB) mechanisms~\cite{bib6, bib7} are applied to decide the set of channels for transmitting on a per-packet basis, thus potentially improving performance. In \cite{bib7}, the following DCB policies were proposed and analyzed:
\begin{itemize}
	\item \textbf{Static Channel Bonding (SCB):} a transmitter is allowed to use the entire set of channels only, thereby limiting the election of any subset of channels. While such a policy may optimize the performance in isolated deployments, it lacks the necessary flexibility to deal with inter-BSS interference.
	\item \textbf{Always-Max (AM):} in this case, the widest combination of channels is picked upon having sensed them free during the backoff procedure. While such a policy seems to properly harness the available spectrum, it has also been shown to generate starvation and other issues as a result of inter-BSS interactions.
	\item \textbf{Probabilistic Uniform (PU):} as an alternative to SCB and AM, PU is introduced to add some randomness in the process of picking free channels, so that any combination is chosen with the same probability. This policy has been shown to improve both SCB and AM in some scenarios in which flow starvation was present due to inter-BSS interactions.
\end{itemize}

To better illustrate the behavior of CB policies, Fig.~\ref{fig:fig3} shows a simplification of the transmission procedure that an STA follows when implementing AM. In particular, CB may be applied over channels 1-4. Based on the AM policy, at time $t_1$, the STA can transmit only over channel 2, which is the only one that is sensed free at the moment of initiating a transmission. Similarly, in $t_2$, both channels 1 and 2 are found free, so the transmission is performed over those two channels. Finally, provided that the entire spectrum is free, a transmission over channels 1-4 is performed at $t_3$. Notice that, if applying SCB, the STA would have not been able to transmit until $t_3$. Alternatively, regarding PU, any combination of free channels could have been selected in $t_2$ and $t_3$. 

\startfigure
\includegraphics[width=.8\columnwidth]{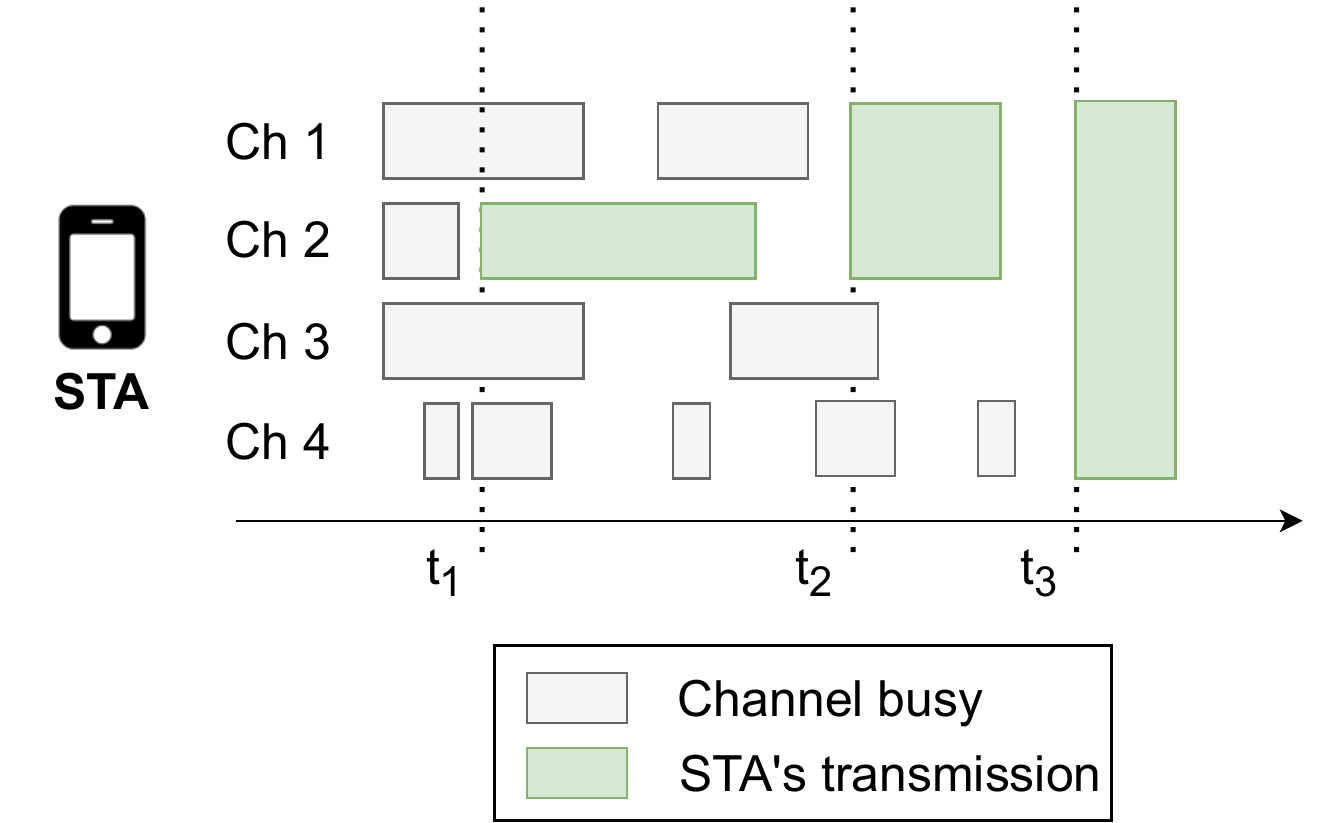}
\caption{Dynamic channel selection for transmitting when applying DCB.}\label{fig:fig3} 
\end{figure}

Through the analysis conducted in~\cite{bib7}, it was shown that the right channel choice is not always trivial (i.e., selecting the widest channel does not necessarily entail achieving the highest performance). First of all, using wider channels entails spreading the same transmit power over the selected channel width, which can potentially affect the data rate used for the transmission, and therefore the capabilities of the receiver on decoding data successfully. Moreover, the potential gains of DCB in crowded deployments are hindered by the interactions among Wi-Fi devices, which may provoke contention or collisions. The fact is that WLAN deployments are unplanned and operate under Carrier Sense Multiple Access (CSMA). From the perspective of a given transmitter-receiver pair, such lack of coordination leads to uncontrolled interference that can potentially degrade their performance.\ITUpar

Other DCB mechanisms were proposed in~\cite{bib91, bib92, bib93, bib94}, which include collision-detection, carrier sensing adaptation, or traffic load awareness. More recently, ML and game theory have been applied to address CB as an online decision-making problem involving multiple agents~\cite{bib95, bib96}.\ITUpar

In view of the complexity of selecting the best configuration of channels, the proposed problem statement had the goal of shedding light on the potential role of ML in DCB. In particular, it served to gather participants' proposals of ML models able to predict the performance of different CB configurations. This information can be used by a decision-making agent to choose the best configuration of channels before initiating a transmission. Throughput prediction in WLANs has been widely adopted for performance analysis through mathematical models, including the well-known Bianchi model~\cite{bib911}, Continuous-Time Markov Networks (CTMNs)~\cite{bib912}, or stochastic geometry~\cite{bib913}. However, these well-known models lack applicability for online decision-making because they fail to capture important phenomena either on the PHY or the MAC, or they entail a high computational cost. Thus, modeling high-dense complex deployments through these models may be highly inaccurate or simply intractable.\ITUpar

In this regard, we envision ML models to assist the CB decision-making procedure in real-time. The fact is that ML can exploit complex characteristics from data, thus allowing to solve problems that are hard to solve by hand-programming (see, for instance, its success in image recognition). Besides, ML models can be trained offline and then improve their accuracy with measurements acquired online. To the best of our knowledge, this is an under-researched subject. While ML has been applied for predicting aspects related to Wi-Fi networks, such as traffic and location prediction~\cite{bib97,bib98}, it has been barely applied for explicitly predicting their performance. In this context, the work in~\cite{bib910} provided an ML-based framework for Wi-Fi operation, which includes the application of Deep Learning (DL) for waveforms classification, so that WLAN devices can identify the medium as idle, busy, or jamming. Closer in spirit to our work,~\cite{bib99} proposed an ML-based framework for WLANs' performance prediction.\ITUpar

\subsection{Introduction to the Dataset}

To motivate the usage of ML for predicting WLANs' performance, we provide an open dataset\ITUfootnote{The dataset has been made publicly available at \ITUurl{https://zenodo.org/record/4059189}, for the sake of openness.} obtained with the Komondor simulator.\ITUfootnote{\ITUurl{https://github.com/wn-upf/Komondor}, Commit: d330ed9.} Komondor is an open-source IEEE 802.11ax-oriented simulator, which fundamental operation has been validated against ns-3 in~\cite{bib8}. Komondor was conceived to cost-effectively simulate complex next-generation deployments implementing features such as channel bonding or spatial reuse~\cite{bib9}. Besides, it includes ML agents, which allows simulating the behavior of online learning mechanisms to optimize the operation of WLANs during the simulation.\ITUpar

The dataset generated with Komondor has been used for training and validate ML models in the context of the ITU AI for 5G Challenge. The assets provided comprise both training and test data sets corresponding to multiple random WLAN deployments at which different CB configurations are applied. As for training, two separate enterprise-like scenarios, namely, \textit{training1} and \textit{training2}, have been characterized. In each case, a different fixed number of BSSs coexist in the same area, according to users' density.\ITUpar

In \textit{training1}, there are 12 APs, each one with 10 to 20 associated STAs. Regarding \textit{training2}, it contains 8 APs with 5 to 10 STAs associated with each one. For both training scenarios, three different map sizes have been considered (\textit{a}, \textit{b}, and \textit{c}), where STAs are placed randomly. Similarly to training scenarios, the test dataset includes a set of random deployments depicting multiple CB configurations and network densities. In this case, four different scenarios have been considered according to the number of APs (4, 6, 8, and 10 APs). Note, as well, that, for each type of scenario, 100 and 50 random deployments have been generated for training and test, respectively. In all the cases, downlink UDP traffic was generated in a full-buffer manner (i.e., each transmitter has always packets to be delivered). Table~\ref{tab:tab1} summarizes the entire dataset in terms of the simulated deployments.\ITUpar

\starttable
\caption{Summary of the simulated deployments used for generating both training and test datasets.}\label{tab:tab1} 
\begin{small}
\begin{tabular}{|c|c|l|c|c|}
	\hline
	& \textbf{Scenario id} & \textbf{Map width} & \textbf{\# APs} & \textbf{\# STAs} \\ \hline
	\multirow{6}{*}{\textbf{Training}} & training1a & 80 x 60 m & \multirow{3}{*}{12} & \multirow{3}{*}{10-20} \\ \cline{2-3}
	& training1b & 70 x 50 m &  &  \\ \cline{2-3}
	& training1c & 60 x 40 m &  &  \\ \cline{2-5} 
	& training2a & 60 x 40 m & \multirow{3}{*}{8} & \multirow{3}{*}{5-10} \\ \cline{2-3}
	& training2b & 50 x 30 m &  &  \\ \cline{2-3}
	& training2c & 40 x 20 m &  &  \\ \hline
	\multirow{4}{*}{\textbf{Test}} & test1 & \multirow{4}{*}{80 x 60 m} & 4 & \multirow{4}{*}{2-10} \\ \cline{2-2} \cline{4-4}
	& test2 &  & 6 &  \\ \cline{2-2} \cline{4-4}
	& test3 &  & 8 &  \\ \cline{2-2} \cline{4-4}
	& test4 &  & 10 &  \\ \hline
\end{tabular}
\end{small}
\end{table}

With respect to input features, these are included in the files used for simulating each random deployment. In particular, the most relevant information to be used for training ML models is:
\begin{enumerate}
	\item \textbf{Type of node:} indicates whether the node is an AP or an STA.
	\item \textbf{BSS id:} identifier of the BSS to which the node belongs.
	\item \textbf{Node location:} \{x,y,z\} coordinates indicating the position of the node in the map.
	\item \textbf{Primary channel:} channel at which carrier sensing is performed.
	\item \textbf{Channels range:} minimum and maximum channels allowed for bonding.
	\item \textbf{Transmit power:} power used for transmitting frames.
	\item \textbf{Sensitivity threshold:} used for detecting other transmissions and assess the channels' availability.
	\item \textbf{Received Signal Strength Indicator (RSSI):} power detected at receivers from their corresponding transmitters.
	\item \textbf{Inter-BSS interference:} power sensed from other ongoing transmissions.
	\item \textbf{Signal-to-Interference-plus-Noise Ratio (SINR):} average SINR experienced during packet receptions.
\end{enumerate}

Regarding output labels, we provide the throughput obtained by each device during the simulation, being the APs' throughput the aggregate throughput of each BSS (i.e., the sum of all the individual STAs' throughput in a given BSS). Besides, the airtime per AP is provided, which indicates the percentage of time each BSS has occupied each of its assigned channels.

\section{Machine Learning Solutions for Throughput Prediction}
\label{section:solutions}

In this Section, we overview the solutions proposed by the participating teams of PS-013 in ITU AI for 5G Challenge: \textit{ATARI} (University of Antwerp and Universidad de Antioquia), \textit{Ramon Vallés} (Universitat Pompeu Fabra), \textit{STC} (Saudi Telecom), \textit{UC3M NETCOM} (Universidad Carlos III de Madrid), and \textit{Net Intels} (PES University). From these teams, ATARI, Ramon Vallés, and STC succeeded to advance to the Grand Finale, where teams from all the problem statements competed for winning the global challenge~\cite{problems}.

\subsection{ATARI}
Wireless networks can be represented by graphs G=(V, E), where V is the set of nodes, i.e., STAs and APs, and E represent wireless links. Typically, DL approaches deal with graph-structured data by processing the data into simpler structures, e.g., vectors. However, nodes and links in high dense WLAN deployments are characterized by a set of high-dimensional features, thus complicating the graph-like structure of the problem and therefore hindering the application of deep learning.\ITUpar

To overcome the problem of data representation, Graph Neural Networks (GNNs) have been proposed as neural networks that operate on graphs intending to achieve relational reasoning and combinatorial generalization~\cite{bib10}. Accordingly, the wireless interactions between STAs and APs (connectivity, interference, among others) can be easily captured via a graph representation. \ITUpar

Therefore, we select a GNN approach to predict the throughput of the devices in a WLAN. In particular, each deployment is considered as a directed graph where STAs and APs are the graph's nodes. Besides, we define two types of nodes present in the dataset, each one having generic or specific features. For instance, parameters like channel configuration are related to both types of nodes, while SINR is only related to STAs, and airtime is exclusive for APs. Furthermore, the edges are defined based on the type of wireless interaction derived from the dataset. We consider two types of interactions, namely AP-AP interactions (represented by the interference map), and AP-STA interactions (represented by the RSSI values). For completeness, we define an additional edge feature based on the distance of every transmitter-receiver pair. The features considered for training the proposed GNN are summarized in Table~\ref{tab:tab2}. \ITUpar

\starttable
\caption{Features used by ATARI team to train a GNN.}\label{tab:tab2} 
\begin{small}
\begin{tabular}{|c|c|c|}
	\hline
	& \textbf{Feature} & \textbf{Pre-processing} \\ \hline
	\multirow{7}{*}{\textbf{Node}} & Node type & AP=0, STA=1 \\ \cline{2-3} 
	& Position (x,y) & None \\ \cline{2-3} 
	& Primary channel & \multirow{3}{*}{\begin{tabular}[c]{@{}c@{}}Combined into a \\ categorical variable \\ using one-hot encoding\end{tabular}} \\ \cline{2-2}
	& Min. channel &  \\ \cline{2-2}
	& Max. channel &  \\ \cline{2-3} 
	& SINR & None \\ \cline{2-3} 
	& Airtime & Mean \\ \hline
	\multirow{4}{*}{\textbf{Edge}} & Edge type & AP-AP=0, AP-STA=1 \\ \cline{2-3} 
	& Distance & Computed from (x,y) \\ \cline{2-3} 
	& RSSI & None \\ \cline{2-3} 
	& Interference & None \\ \hline
\end{tabular}
\end{small}
\end{table}

Concerning the GNN model, we have used an implementation of the Graph Network Block (GNB), as proposed in~\cite{bib11}. A GNB contains three update functions and three aggregation functions where the computation is done from edge to nodes, and then to global parameters. So first, the edge's features are updated and aggregated into the node features, then the node features are updated having into account the vicinity within depth/range defined by the number of GNBs, and lastly, the global parameters are updated according to the state of the nodes. Our model follows a layered approach, similar to DL, where each layer is a GNB. The input of the model is a graph representing the deployment, and the output is the predicted throughput of the devices in that deployment. A general overview of our model's architecture is shown in Fig.~\ref{fig:ATARI-GNN}.\ITUpar

\startfigure
\includegraphics[width=\columnwidth]{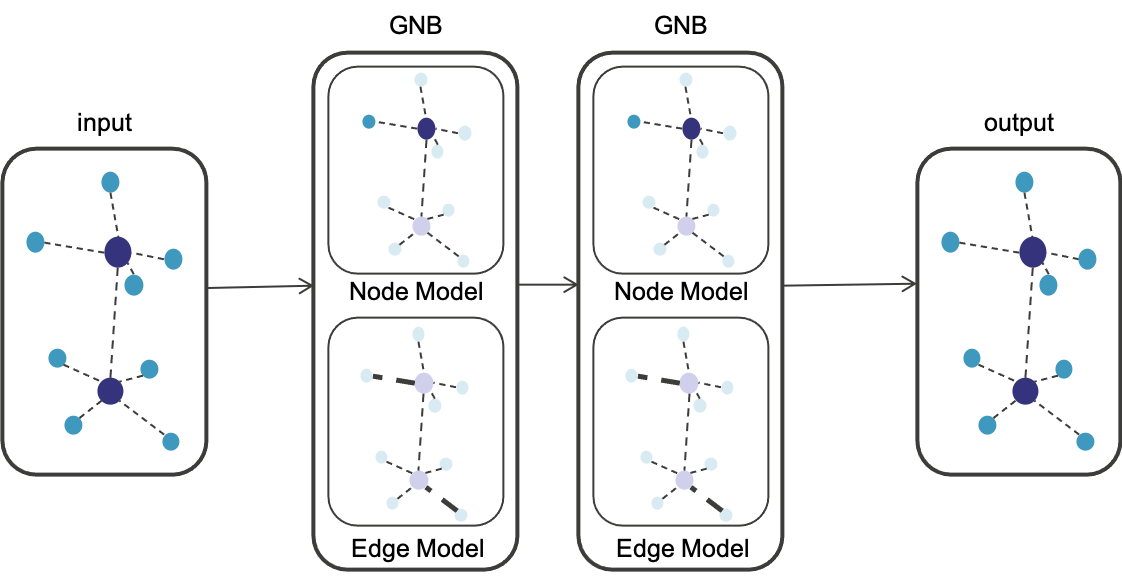}
\caption{GNN model proposed by ATARI.}\label{fig:ATARI-GNN} 
\end{figure}

The implementation of the GNB is referenced as a MetaLayer in PyTorch Geometric~\cite{bib12}, a geometric deep learning extension library for PyTorch. We defined an edge model that uses two dense layers using Rectified Linear Unit (ReLU) as an activation function in a typical Multi-Layer Perceptron (MLP) configuration. A node model is also defined by two MLPs, one for aggregating the edge features into the node features and the second to update nodes' states based on their neighbors. Note that, following the formal definition of a GNB in~\cite{bib11}, global parameters were not used. Our implementation is available in GitHub~\cite{bib13}. \ITUpar

To train the model, we have considered splitting the provided dataset (80\% for training and 20\% for validation). As each deployment is considered to be represented by a graph, 480 graphs have been used for training, and 120 for validating purposes. The loss function considered to assess the performance of our model is the Root Mean Squared Error (RMSE). The error obtained across all the predictions is used to compare the accuracy of our model's predictions to the actual results. However, initial results showed that the model mostly focused on predicting the throughput of the APs, given that the error is minimized on large values. Therefore, we proposed a masked loss provided that AP’s throughput should be equal to the sum of the associated  STAs’ throughput. The RMSE is calculated using the STAs' predicted and the APs' computed throughput.\ITUpar

\subsection{Ramon Vallés}
To address the throughput prediction problem in CB-compliant WLANs, we propose a deep neural network where the information of each BSS is processed independently, thus following the idea of Multi-Layer Perceptron (MLP)~\cite{multilayer_perceptron}. More specifically, the proposed model is a feed-forward deep learning algorithm implemented in python with the support of the PyTorch libraries.\ITUfootnote{The code used to implement the method proposed by Ramon Vallés is open-access~\cite{ramon_github}.} \ITUpar

Our model aims to predict the aggregate throughput of each BSS, rather than the individual throughput at STAs. The fact is that predicting the throughput per STA is very challenging because of the dynamic channel bonding policy used in complex scenarios, which contributes to generating multiple interactions among nodes that cannot be captured at a glance. Accordingly, to derive an overall representation of each BSS, the features from individual STAs are preprocessed so that we consider only their global distribution (mean, and standard deviation). To do so, our model is divided into three main blocks performing different tasks:
\begin{enumerate}
	\item \textbf{Signal quality:} The first block is meant to abstract the interactions among APs and STAs in the RF domain. To achieve this, we consider two separate layers, which process the RSSI, SINR, distance among nodes, and SINR. A parametric Rectified Linear Unit (PReLU) activation function is used together with 1-dimensional batch normalization. 
	\item \textbf{AP bandwidth:} The second block analyzes the available bandwidth of the APs, and it consists of a single linear layer, which receives as input a vector with the corresponding airtime for all the available channels. This layer outputs a 3-dimensional array and is activated with a PReLU function.
	\item \textbf{Output:} Finally, the last block takes the output from both Signal quality and AP bandwidth blocks and computes the final prediction value. For this final layer, we employ a simple ReLU (instead of a PReLU) to avoid negative throughput predictions.
\end{enumerate}

With this structure, we have built a much more efficient model than if we had used a fully-connected NN with all the STA features. Notice that the proposed model needs far fewer neurons (and thus, less computational force) to capture the most relevant information of each scenario.  In our opinion, an excess of neurons in such a complex scenario would result in overfitting, thus making the model less accurate for predicting the performance of new deployments.\ITUpar

As for training the MLP model, we have considered the following key features: the type of device (AP or STA), its location, the minimum, and maximum channel through which it is allowed to transmit, the RSSI, the SINR, and the airtime. The training was performed using 80\% of the dataset (keeping the 20\% left for validation), following an evaluation criterion based on the RMSE of the predicted throughput with respect to the real value. The Adam optimizer has been used to optimize the training process, which is straightforward to implement, computationally efficient, and memory-efficient. For the training phase, several experiments had been made by modulating the hyper-parameters. The best results were achieved with a learning rate of 0,025 and a total of 700 epochs.\ITUpar

\subsection{STC}
Our proposal includes popular ML regression algorithms such as MLP, Support Vector Machine (SVM), Random Forest, and eXtreme Gradient Boosting (XGboost). These algorithms are backed by rich research, known to do well on regression problems, ease of implementation and deployment, which are important characteristics from the business perspective.\ITUpar

From all the abovementioned methods, XGBoost~\cite{xgboost} was selected for competing in the challenge because it offered the highest performance on both the training and validation stages compared with the other models. XGboost is a gradient boosting framework available for multiple platforms, thus providing high portability.\ITUpar

To train our model, we have analyzed the variance of the variables in the dataset, and discarded the features with low or moderate variability. Some of the considered features are the position of nodes, the primary channel, the distance among nodes, or the power received by neighboring devices. To pre-process the selected features, we have applied Yeo-Johnson transformation~\cite{yeo} and normalization. These steps were done for predictors to improve their utility in the models.\ITUpar

The training dataset was split into training and validation, so that we could train the model on the whole training dataset using a 10-fold cross-validation procedure. Further, we used 100 and 300 combinations, respectively, with 10-fold cross-validation.\ITUpar

As for the hyperparameter setting (e.g., max depth or minimum child weight), we tuned hyperparameters using a grid search. More specifically, the hyperparameter values were set using Latin Hypercube Sampling, while their maximum and lower range value for each hyperparameter were mostly pre-determined using default values from tidymodels~\cite{kuhn}.\ITUpar

Finally, significant efforts have been put to deploy our model. In particular, we have used docker to make our model easy to (re)train and deploy. All the code and documentation has been made publicly available~\cite{github_stc}.\ITUpar

\subsection{UC3M NETCOM}

We formulate throughput forecasting in WLANs as a linear regressing problem, which assumes a linear relationship between the features and label(s) of a given data set \{$y_i$, $x_{i,1}$, $x_{i,2}$, ..., $x_{i,N}$\}, and a set of unknown parameters $w$ to be learned (being $w_0$ the bias).\ITUpar

Our solution (named \textit{Gossip}) is based on a linear regression method, and it aims to predict the throughput of an STA in a given WLAN deployment where CB is applied~\cite{github_netcom}. Based on STAs' individual throughput, we derive the performance of each AP by aggregating the values of their associated STAs. In particular, Gossip derives the unknown bias and weight parameters $w$ by (i) processing the WLAN dataset, and (ii) applying a neural network to perform regression. \ITUpar

As for the processing part, Gossip takes the input features generated by the Komondor simulator, and selects/generates the most relevant ones: the position of the STA, the AP to which the STA is associated, the RSSI, the SINR, the set of nodes using the same primary channel, and the set of allowed channels. After processing the features, each STA of every deployment is characterized by a feature vector ($x_i$,…,$x_{21+3k}$ ), with $k$ denoting the number of wireless channels. Note, as well, that the entire dataset is considered for training, thus combing STAs from different deployments. The rationale is that features such as the number of neighbors in the primary channel, the SINR, and the interference should differentiate STAs from different deployments. \ITUpar 

When it comes to the regression problem, Gossip uses a feed-forward neural network with four layers. The input layers pass the input features to two fully connected layers of neurons with a ReLU activation unit. Finally, a single neuron receives the output of the hidden layers and generates the prediction of the throughput. As a remark, the last neuron has a linear activation. It is important to remark that the proposed neural network is mostly meant to tackle linear regression problems. Nevertheless, even if the throughput prediction problem for WLANs is not linear, we expect our model to properly identify local minimum/maximum points that allow providing reasonably prediction results.\ITUpar

To train the proposed neural network, we have used the RMSprop gradient descend method, considering the Mean Squared Error (MSE) as a loss function. Moreover, 50 training episodes and a batch size of 50 STAs have been considered. Thanks to Gossip design, the training dataset is populated with every STA of every deployment present among all scenarios.

\subsection{Net Intels}
To address the objective of predicting the throughput of APs and STAs in typical dense environments, we explore a set of popular regression techniques. With the help of these techniques, we aim to build complex mathematical relationships among features and labels from the dataset, so that performance of WLANs can be predicted at unseen deployments. In particular, we propose using the following techniques:\ITUfootnote{The code used to implement all the proposed methods by Net Intels is available in Github~\cite{netintels_github}.}
 \begin{enumerate}
	\item \textbf{Artificial Neural Network (ANN):} ANN method is selected chiefly due to its potential and versatility to model nonlinear and complex relationships in OBSS data elegantly. The proposed ANN is built using Tensorflow and Keras libraries in python~\cite{keras}. The NN model is designed with one input layer, 7 hidden layers, and 1 output layer (see Fig.~\ref{fig:net_intels_1}). The ReLU function is employed to activate hidden layers. In each of the first six hidden layers, there are 1024 nodes. For the seventh hidden layer, there are 512 nodes. The model is trained using Adam optimizer. The batch size and number of epochs for training, after multiple trials, were set to 250 and 1000 respectively.
	\item \textbf{K-Nearest Neighbor (KNN) regression:} For OBSSs involving several AP-STA combinations, the dynamics of interrelations between entities of OBSS rely predominantly on the relative positioning of AP/STAs. To abstract such complexity in a cost-effective manner, KNN is selected, which is characterized by its simplicity, speed, and protection against high variance and bias. The KNN model is built using the Scikit Learn library in python~\cite{scikit}. The inbuilt \textit{KNearestRegressor} function is directly used, which neighbor number is fixed to 10. The algorithm for structuring the k-dimensional space of the dataset (Ball Tree, KDTree, or Brute Force) is automatically selected based on input values.
	\item \textbf{Random Forest Regression:} Motivated by the fact that the interrelationship between the features is non-linear, we propose dividing the dataset dimensional space into smaller subspaces. To generalize the data and for better feature importance, an ensemble of trees forming a Random forest is used. Random forest mechanisms are useful to reduce the spread and diversion of predictions. The proposed Random Forest regression is built using Scikit Learn, an ensemble module of the Sklearn library. The default number of trees was set to 100, which split is performed according to the mean squared error function. The maximum depth of the tree is set to 10.
\end{enumerate}

\startfigure
\includegraphics[width=\columnwidth]{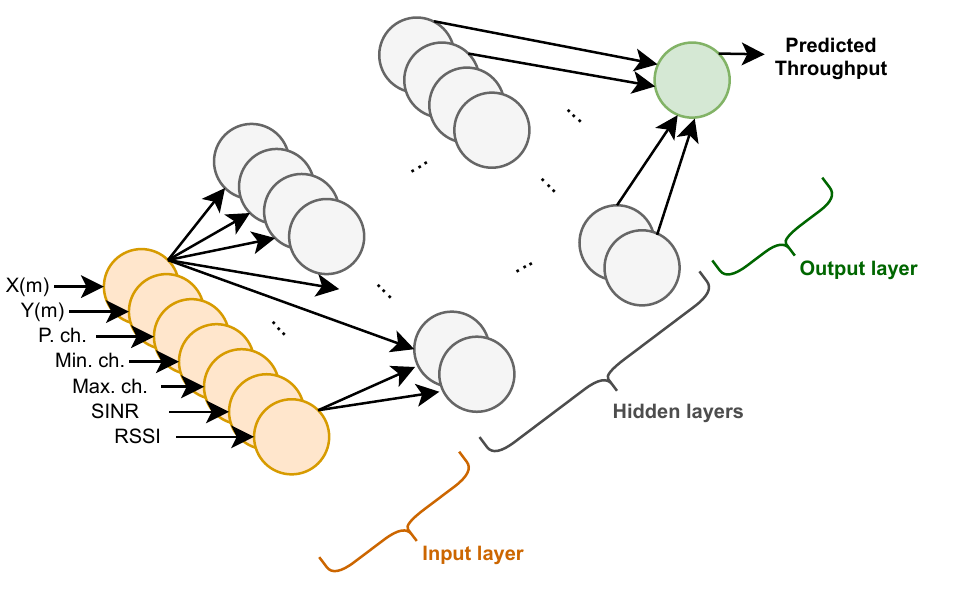}
\caption{Net Intels' ANN architecture.}\label{fig:net_intels_1} 
\end{figure}

For all the proposed methods, we have first preprocessed the dataset comprising six hundred different random deployments. In particular, static features such as the Contention Window (CW) were not included for training purposes. As for the rest of the features, the P-values of each feature were considered to identify the ones with higher significance. To firm up the choice of features to be selected for training, we have used the ones with a higher correlation degree (see Fig.~\ref{fig:heatmap_netintels}).\ITUpar

\startfigure
\includegraphics[width=\columnwidth]{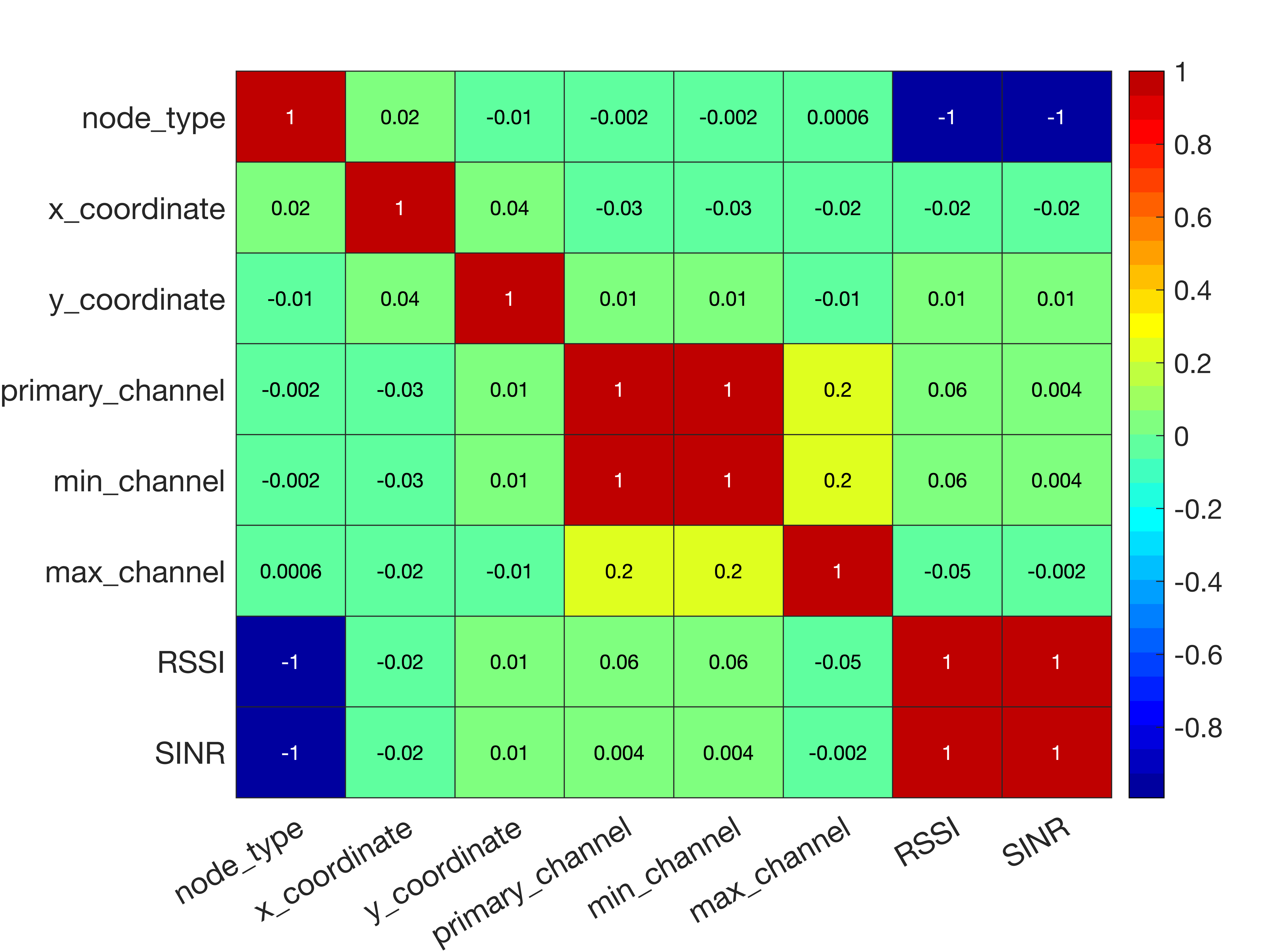}
\caption{Correlation among input features.}\label{fig:heatmap_netintels} 
\end{figure}

Based on this analysis, we decided to include the following features: X and Y coordinates, primary channel, minimum and maximum channel allowed, SINR, and RSSI. The features data is standard scaled before being fed into the regression models. Only data for STAs (stations) are considered for training and the throughput values of STAs are predicted using the models. The sum of the throughputs of STAs gives the throughput of the respective AP. For training purposes of all the three methods, the data is split (80\% for training and 20\% for validation).\ITUpar

\section{Performance Evaluation}
\label{section:performance}

In this Section, we show the results obtained by the participants' models presented in Section~\ref{section:solutions}. In the context of the ITU AI for 5G Challenge, a test dataset was released to assess the performance of each model, without revealing the actual throughput obtained through simulations. Participants were asked to predict the performance in Mbps of each BSS in the test scenarios.\ITUpar

The test dataset consists of random deployments with different characteristics than the ones provided in the training dataset, ranging from low to high density in terms of the number of BSSs and users. In total, test scenarios consist in 200 random deployments containing 1.400 BSSs and up to 8.431 STAs (randomly generated). To assess the participants' model accuracy, we focused on the throughput of the BSSs in each deployment (i.e.,~the throughput of each AP). Specifically, we used both the RMSE and the Mean Absolute Error (MAE) as reference performance metrics. Accordingly, Fig.~\ref{fig:mean_results_all} shows the MAE in Mbps obtained by each team in each type of test scenario.

\startfigure
\includegraphics[width=\columnwidth]{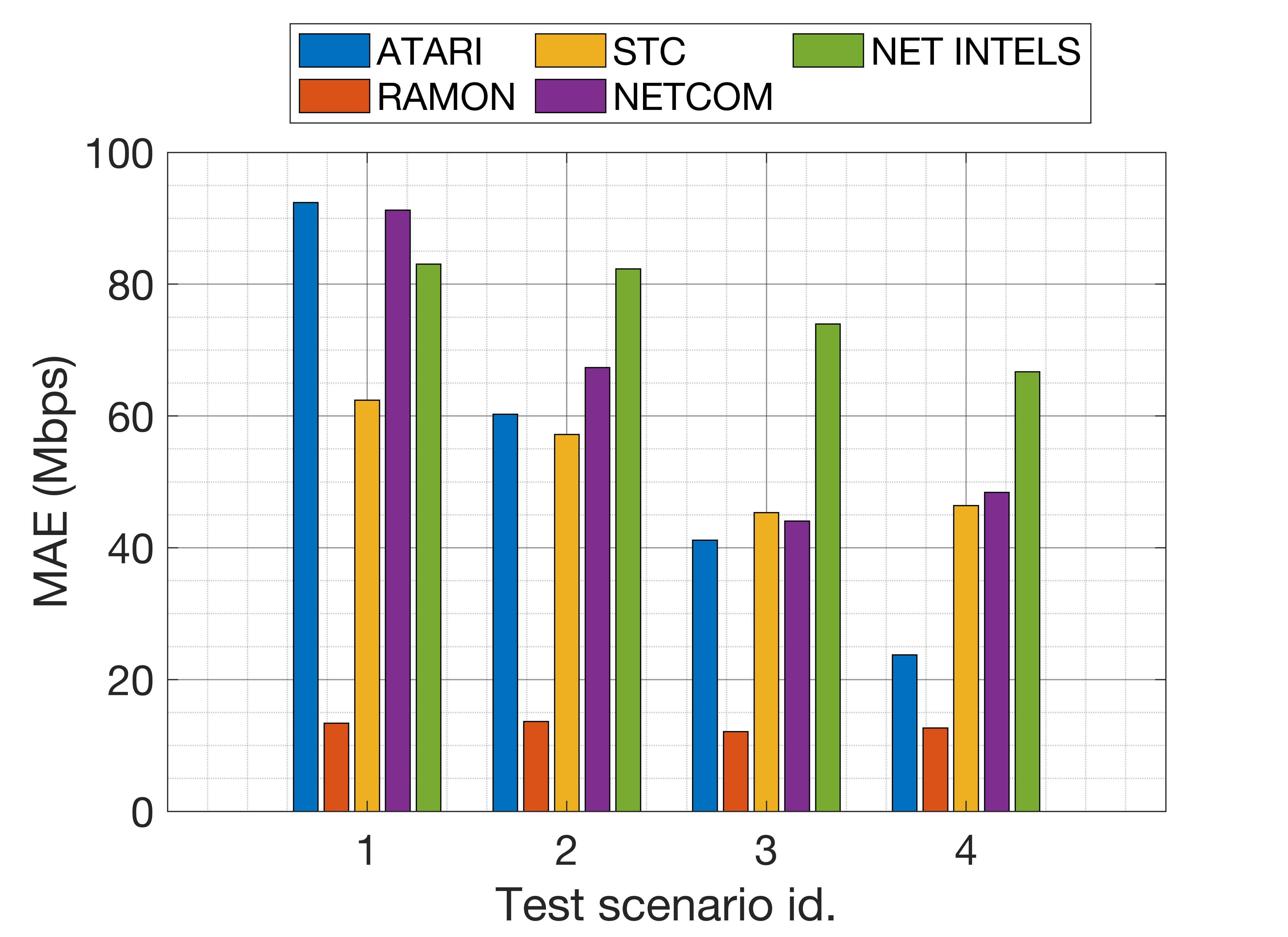}
\caption{Mean absolute error obtained by each team, for each of the test scenarios of the dataset. The aggregate throughput of all the BSSs is considered.}\label{fig:mean_results_all} 
\end{figure}

As shown, for the aggregate BSS performance, most of the models offer low accuracy for the less dense scenarios (namely, \textit{test1} and \textit{test2}), whereas higher accuracy is achieved for the densest deployments (namely, \textit{test3} and \textit{test4}). The fact is that denser deployments are much more similar to the training scenarios than the sparser ones. As a result, models behave pessimistically in low-density deployments by assuming lower performance even if interference is low. As an exception, we find the model provided by Ramon Vallés, a feed-forward neural network with three blocks. The main difference of this model with respect to the others is that it separates the features related to signal quality and interference, and processes them apart from the rest. As a result, it is able to generalize well, even for new deployments with characteristics unseen in the training phase.\ITUpar

Although the prediction error is high for some test scenarios, it is important to remark that the performance of WLANs applying CB can be up to a few hundreds of Mbps (especially in sparse scenarios with low competition). To better illustrate the accuracy of the proposed models, we now show the prediction results obtained on a per-STA basis. Notice that the following results correspond to the solutions provided by three teams (\textit{ATARI}, \textit{STC}, and \textit{Net Intels}), which solution was based on predicting the throughput of STAs, and providing the aggregate performance afterward. Note, as well, that the target of the challenge was predicting the aggregate throughput in each BSS. In particular, Fig.~\ref{fig:histogram_atari_stc_net} shows the histogram of the individual throughput predictions at STAs obtained across all the random test deployments.

\startfigure
\includegraphics[width=\columnwidth]{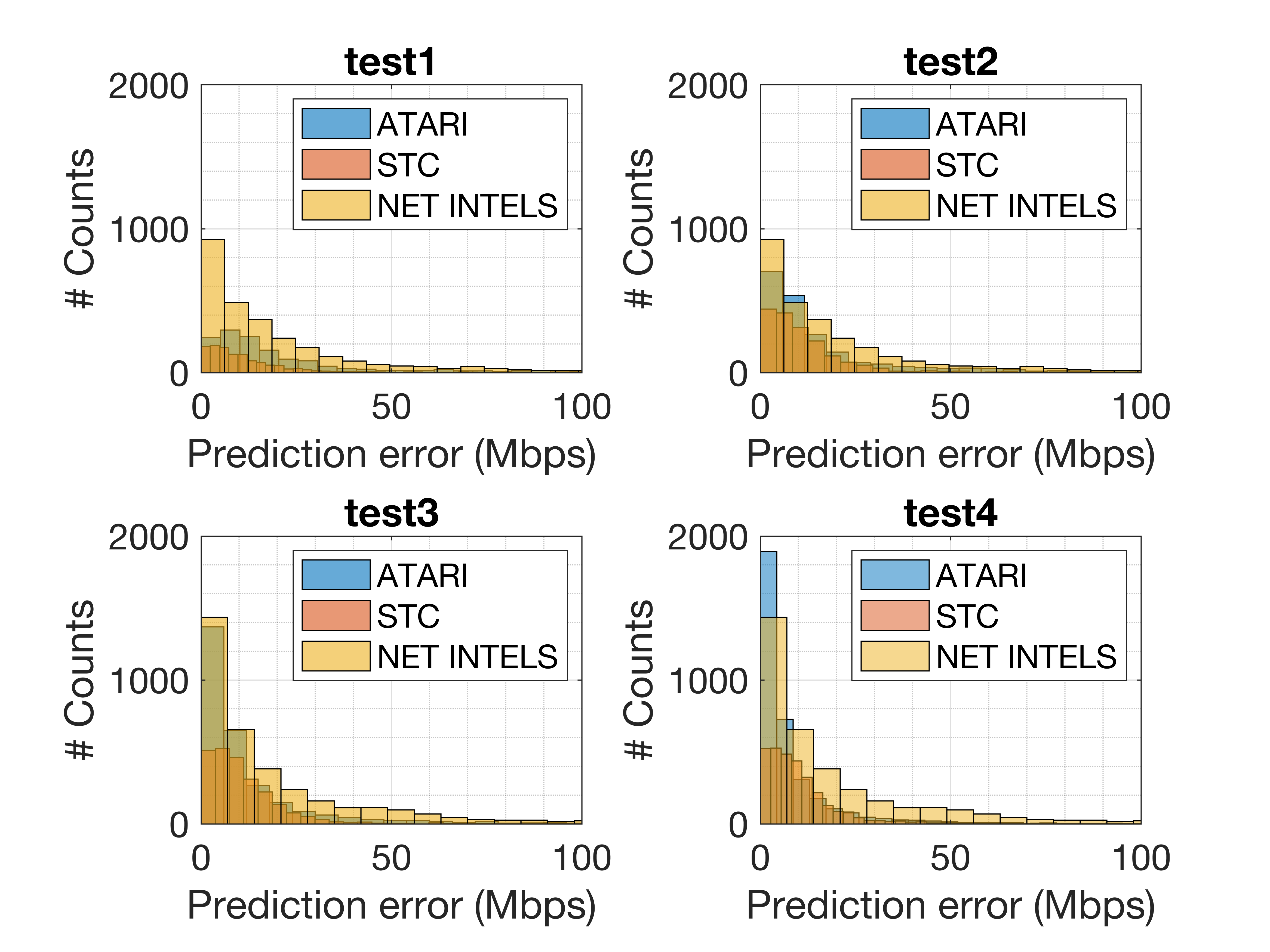}
\caption{Histogram of the per-STA prediction error achieved by ATARI, STC, and NET INTELS.}
\label{fig:histogram_atari_stc_net} 
\end{figure}

As shown, the proposed ML mechanisms provide general accurate predictions; most of the error values are in the range of 0 to 10 Mbps. This means that, even in the presence of outliers, the predictions provided by the ML models are suitable for a significant percentage of the deployments. The accuracy of the different models proposed by ATARI, STC, and NET INTELS can be further observed in Table~\ref{tab:tab3}, which shows the percentage of the throughput predictions for STAs achieving an error below 10 Mbps.

\starttable
\caption{Percentage of per-STA predictions achieving <10 Mbps error. Information is provided for ATARI, STC, and NET INTELS results.}\label{tab:tab3} 
\begin{small}
	\begin{tabular}{|c|c|c|c|c|}
		\hline
		\textbf{} & \textbf{test1} & \textbf{test2} & \textbf{test3} & \textbf{test4} \\ \hline
		\textbf{ATARI} & 36.97\% & 55.81\% & 67.01\% & 77.40\% \\ \hline
		\textbf{STC} & 55.97\% & 56.27\% & 56.74\% & 60.67\% \\ \hline
		\textbf{NET INTELS} & 38.09\% & 42.15\% & 44.01\% & 49.77\% \\ \hline
	\end{tabular}
\end{small}
\end{table}

For completeness, Fig.~\ref{fig:boxplot_tpt_stas} shows the actual throughput achieved by STAs in all the test scenarios. As shown, the median is around 20 Mbps, but maximum values of up to 40 Mbps are also likely. Besides, several outliers were noticed, leading to up to 50 Mbps in some STAs.

\startfigure
\includegraphics[width=\columnwidth]{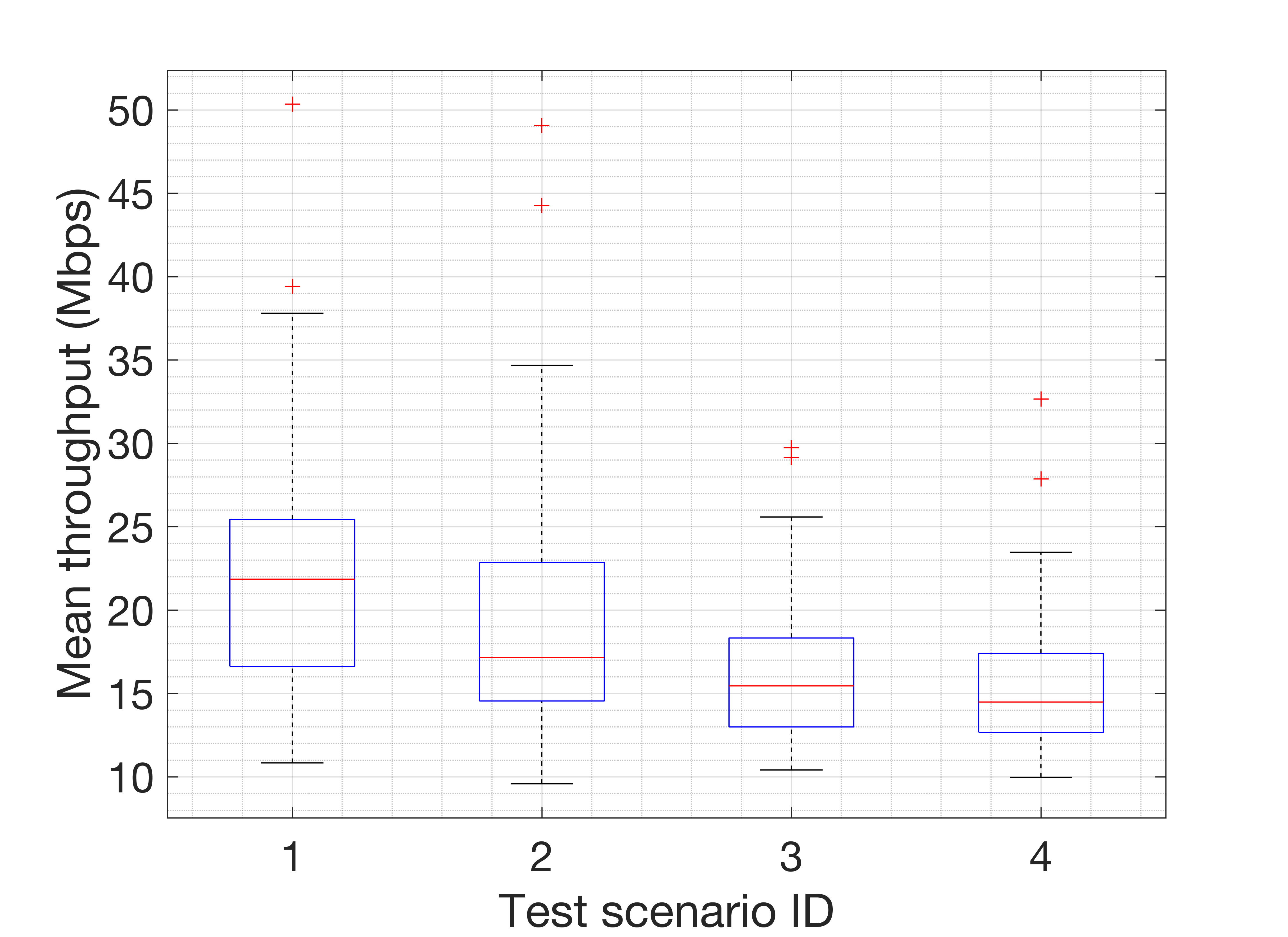}
\caption{Boxplot of the mean throughput achieved by STAs for each test scenario.}
\label{fig:boxplot_tpt_stas} 
\end{figure}

\section{Discussion}
\label{section:conclusions}

\subsection{Contributions}
With contributions from participants around the globe, the ITU AI for 5G Challenge has unprecedentedly established a platform for addressing important problems in communications through ML. As for the performance prediction problem in CB WLANs (referred to as PS-013), the challenge has allowed us to glimpse the potential of ML models for addressing it.\ITUpar

This paper provides a compendium of ML models proposed for throughput prediction in WLANs, including popular models such as neural networks, linear regression, or random forests. In particular, we have overviewed the proposed models and analyzed their performance in the context of the challenge. By opening the dataset used during the competition, we encourage the development of mechanisms that improve the baseline performance shown by the ML models presented in this work.\ITUpar

\subsection{Lessons Learned}

From the performance evaluation done in this paper, we have drawn the following conclusions:
\begin{enumerate}
	\item First, even if the dataset was not particularly big,\ITUfootnote{Youtube-8M Dataset (\ITUurl{http://research.google.com/youtube8m/}) consists of 350.000 hours of video, while only six hundred different random deployments have been used in this paper for training purposes.} some of the proposed ML models achieved good results. This is quite a positive result since it opens the door to ML models that can be (re)trained fast, thus becoming suitable for (near)real-time solutions.
	\item Second, most of the proposed DL-based models have shown higher accuracy for the denser and more complex deployments (which more closely match the training scenarios) than for the sparser ones. While capturing complex situations is quite a positive result, the pitfalls observed in simpler deployments also suggest that out-of-the-box DL methods may fail at capturing the relationship between interference and performance of WLANs. In this regard, well-known models characterizing WLANs (e.g., SINR-based models~\cite{sinrbased}) can potentially be incorporated into the ML operation for the sake of improving accuracy, thus leading to hybrid model-based and data-driven mechanisms.
	\item Third, and related to the previous point, GNNs have been shown to be particularly useful to capture the complex interactions among devices in WLANs, both in terms of interference and neighboring activity. In particular, we have realized the importance of pre-processing the dataset in order to obtain accurate prediction results. Deriving information specific to the problem (i.e., signal quality, interference) has turned out to be essential for the sake of generalization.
	\item Finally, we remark the importance of cost-effectively predicting the performance in WLANs, which may open the door to novel mechanisms using these predictions as heuristics for online optimization. The incorporation of these kinds of models to WLANs is expected to be enabled by ML-aware architectural solutions~\cite{architecture}.
\end{enumerate}

\subsection{The Role of Network Simulators in ML-aware Communications}
The availability of data for training is key for the success of ML application to future 5G/6G networks. Given the current limitations in acquiring data from real networks, simulators emerge as a practical solution to generate complementary synthetic data for training ML models. The fact is that data may be scarce because, among other reasons, measurement campaigns are costly, data from networks involves privacy concerns, or data tenants are not willing to share their data.\ITUpar

Through the problem statement discussed in this paper, we have contributed to showcase how synthetic data can be used to train ML models for networks. A notorious advantage is that network simulators allow characterizing complex deployments, sometimes representing unknown situations, so they help to train and validate ML models. \ITUpar

Beyond generating data for training, network simulators are envisioned to serve as secure platforms for testing, training, and evaluating ML models before being applied to operative networks~\cite{simulators}. In consequence, we foresee the adoption of simulators into future ML-aware networks as a key milestone for enhancing both reliability and trustworthiness in ML mechanisms.


\section*{Acknowledgement}
\label{sec:ackn}
Representative members of each participating team has been invited to co-author this paper. Nevertheless, the same credit goes to the rest of the participants of PS-013 in the ITU AI for 5G Challenge: Miguel Camelo, Natalia Gaviria, Mohammad Abid, Ayman M. Aloshan, Faisal Alomar, Khaled M. Sahari, Megha G Kulkarni, and Vishalsagar Udupi. We would also like to thank enormously everyone that made possible the ITU AI for 5G Challenge, with special mention to Vishnu Ram OV, Reinhard Scholl, and Thomas Basikolo.

This work has been partially supported by grants WINDMAL PGC2018-099959-B-I00 (MCIU/AEI/FEDER,UE), and 2017-SGR-11888.



\newpage
\section*{Authors}
\label{sec:auth}

\startfigure\includegraphics[width=0.4\columnwidth]{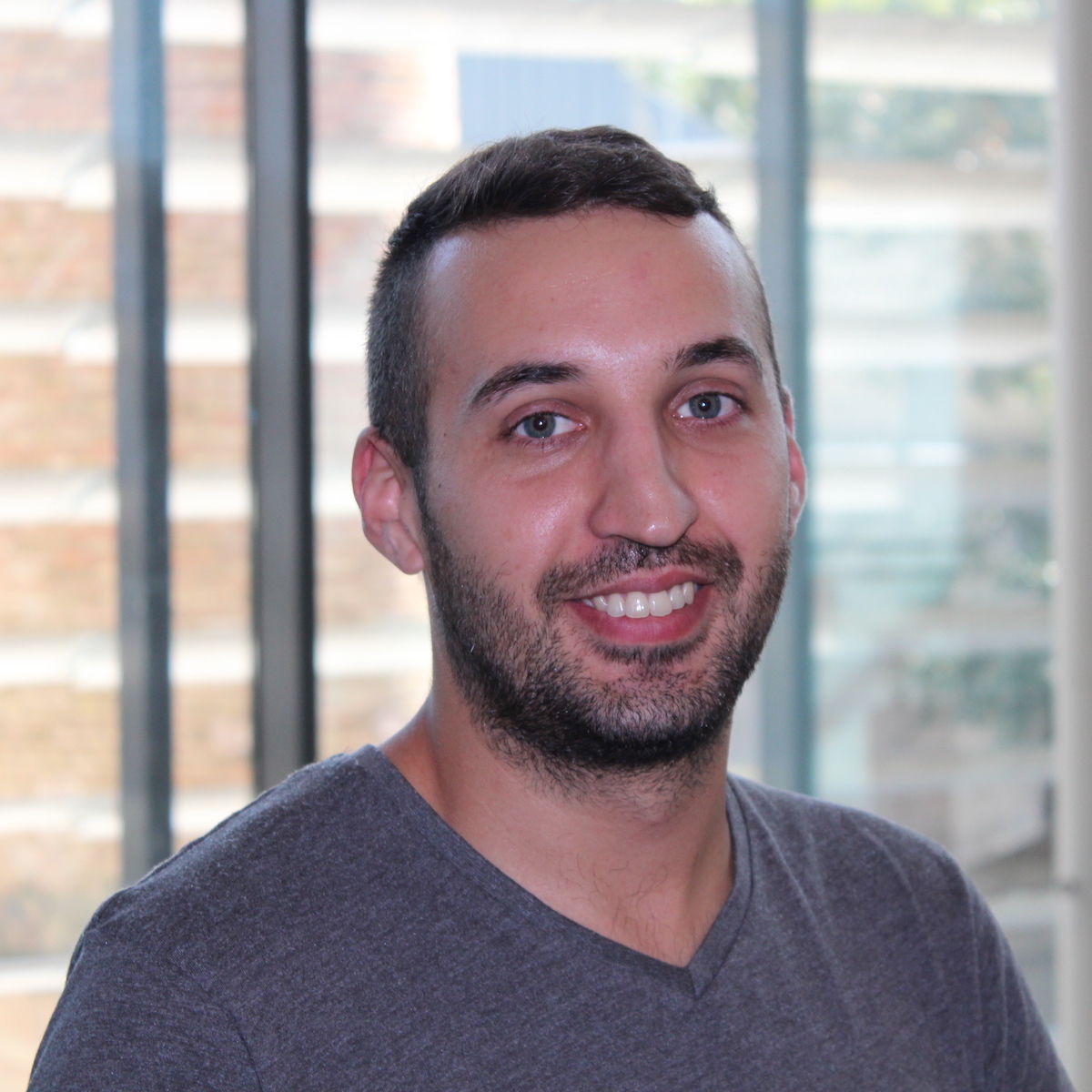} 
\end{figure}\textbf{Francesc Wilhelmi} holds a Ph.D. in Information and Communication Technologies (2020) from Universitat Pompeu Fabra (UPF). He is currently a postdoctoral researcher in the Mobile Networks department at Centre Tecnològic de Telecomunicacions de Catalunya (CTTC) and a teaching assistant at UPF and at Universitat Oberta de Catalunya (UOC).
\ITUpar

\startfigure\includegraphics[width=0.4\columnwidth]{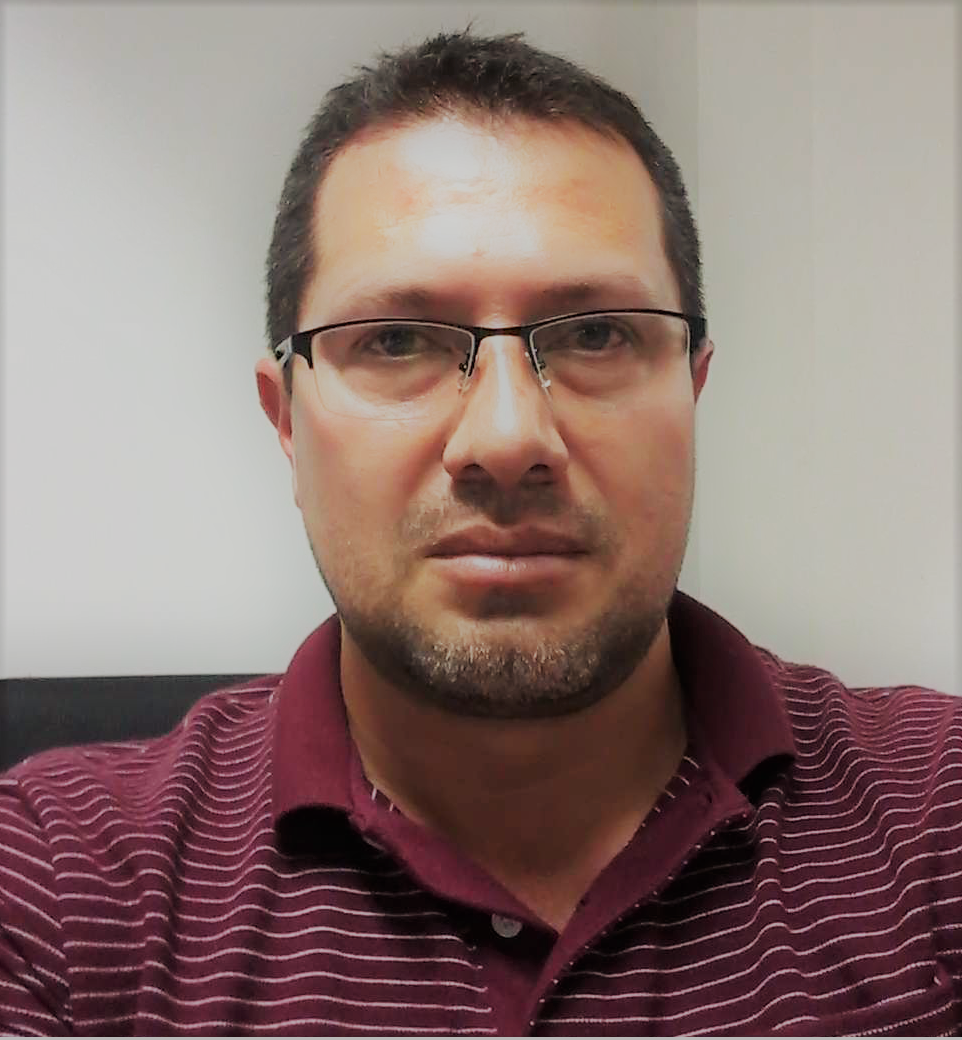} 
\end{figure}\textbf{David Góez} is a Telecommunications Engineer from the ITM Metropolitan Technological Institute, a Master in Automation and Industrial Control from the ITM, a PhD student in Electronic and Computer Engineering at the University of Antioquia. His research focuses on radiocommunication systems whose signal processing blocks can be built with Machine Learning.
 \ITUpar\ITUpar\ITUpar
 
\startfigure\includegraphics[width=0.4\columnwidth]{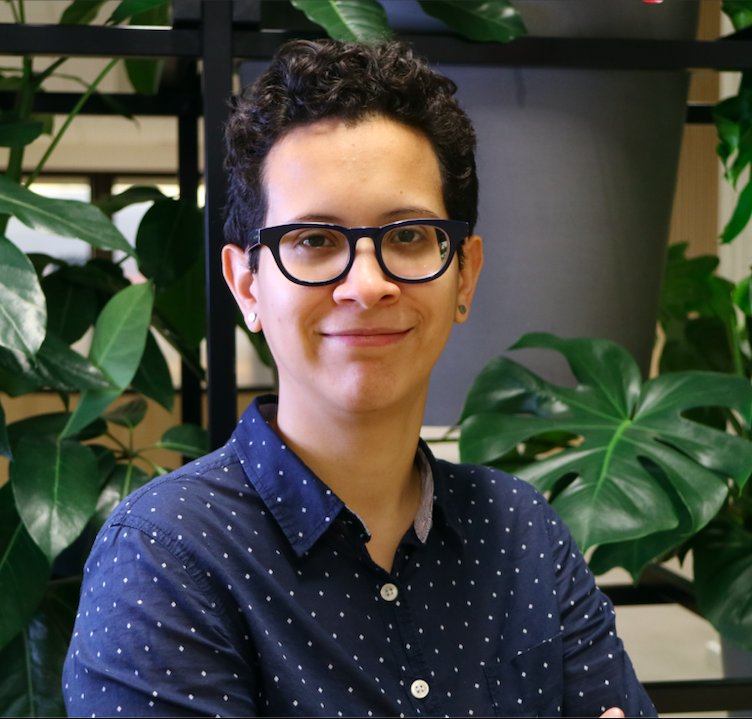} 
\end{figure}\textbf{Paola Soto} is currently pursuing her Ph.D. degree at University of Antwerp - imec, Belgium. Her doctoral research investigates machine learning techniques applied to network management. She received the B.S. degree in Electronics and the M.Sc. degree in telecommunications engineering from the University of Antioquia, Medellín, Colombia, in 2014 and 2018, respectively. Her main research interests include artificial intelligence,as
 machine learning, network management, and resource allocation algorithms.
 
 \startfigure\includegraphics[width=0.4\columnwidth]{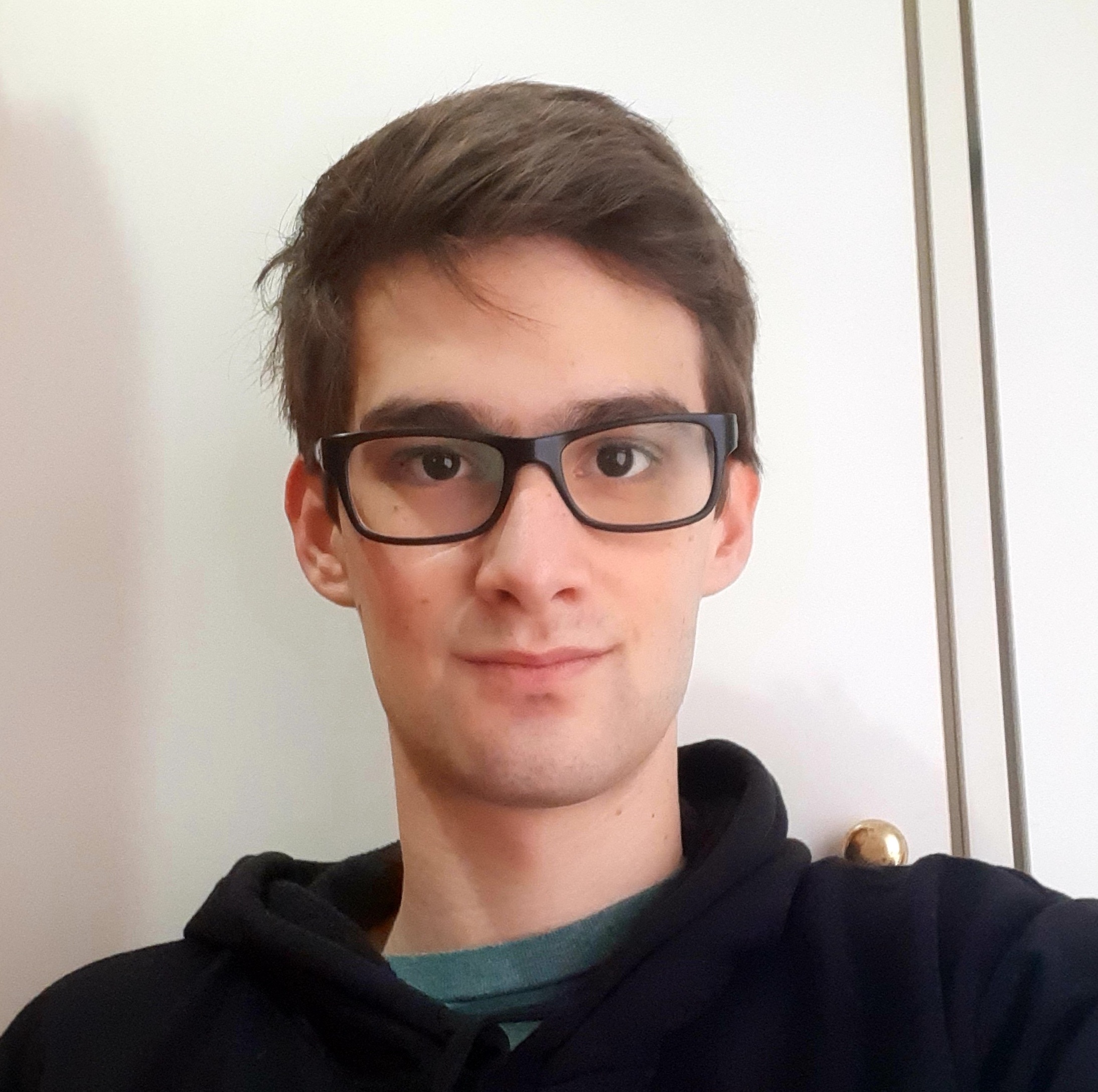} 
\end{figure}\textbf{Ramon Vallés} is currently pursuing a B.Sc. in Computer Science at Universitat Pompeu Fabra (UPF). His research interests are Artificial Intelligence and Machine Learning, Networks, and Network Security. 

\startfigure\includegraphics[width=0.4\columnwidth]{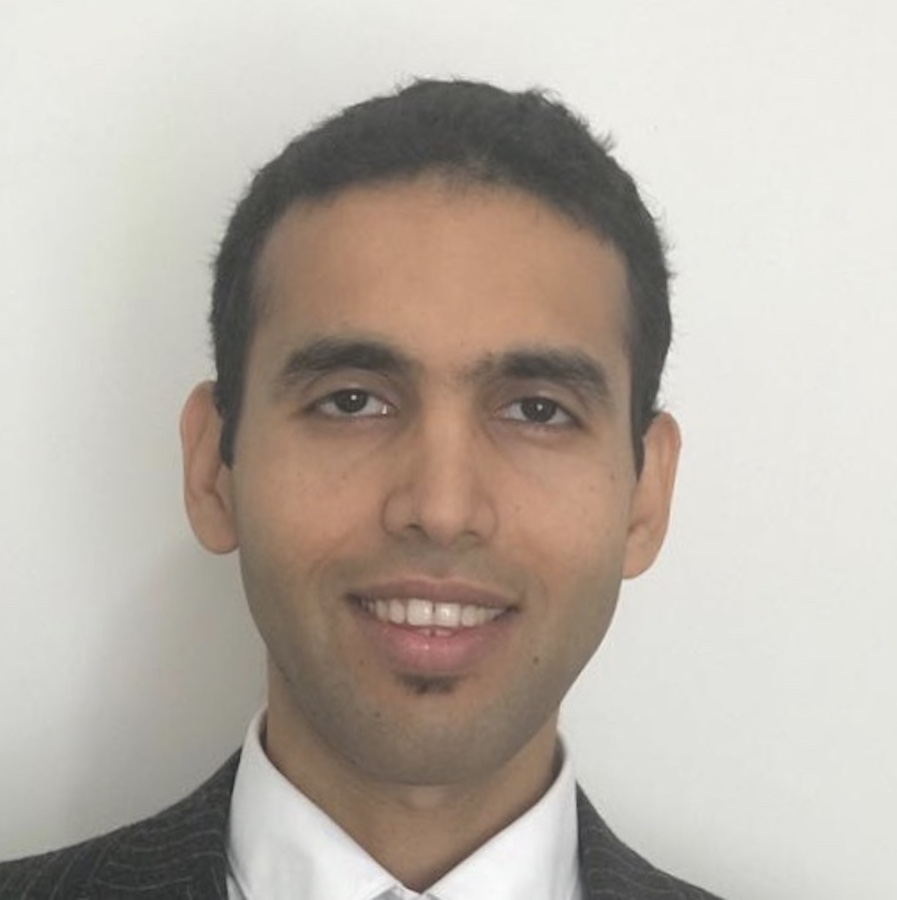} \end{figure}
\textbf{Mohammad Alfaifi} holds an M.Sc. degree in management studies from the University of Waikato (New Zealand). He is currently working as Data Mining Supervisor at Saudi Telecom (STC). 

\startfigure\includegraphics[width=0.4\columnwidth]{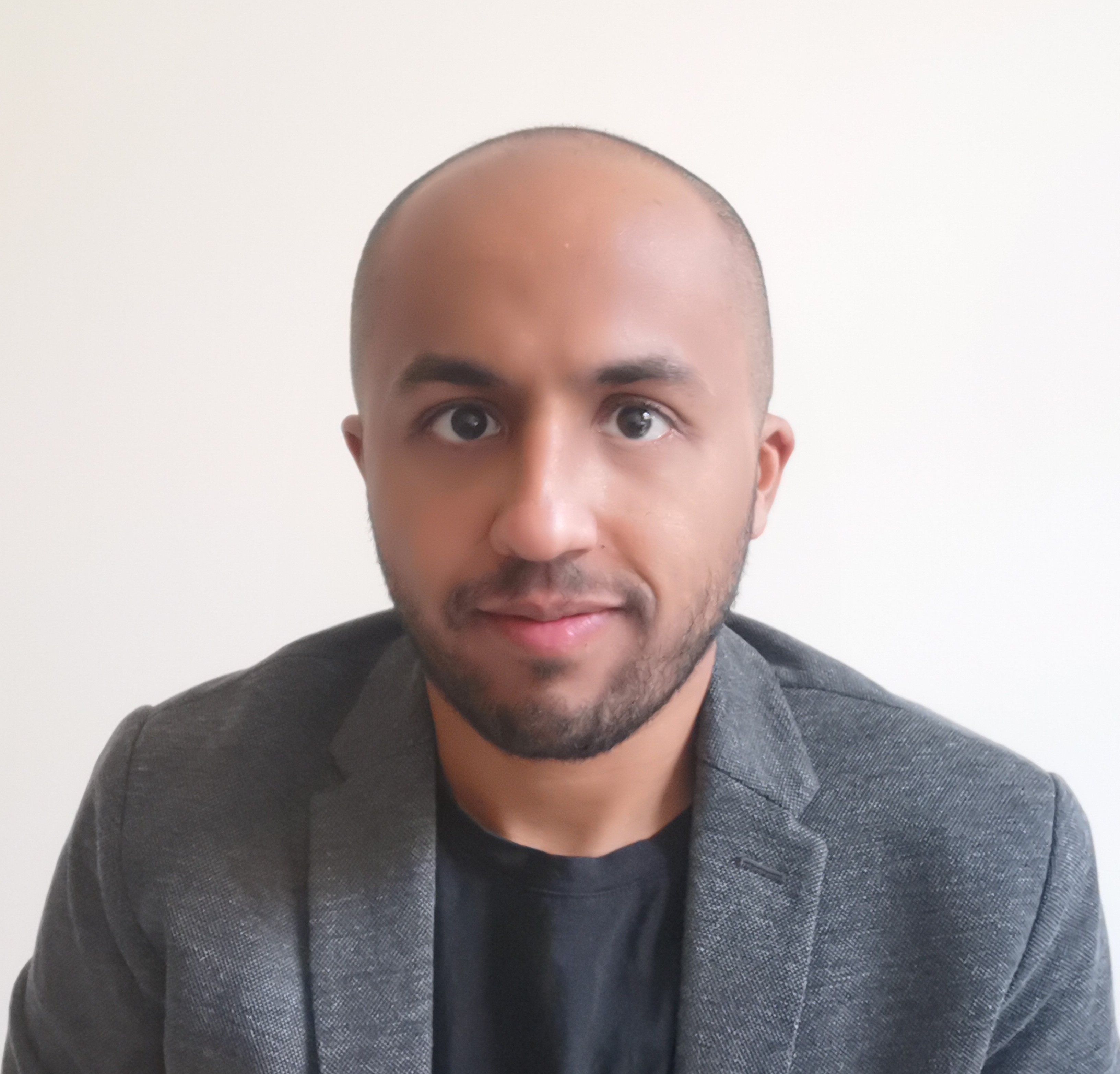} \end{figure}
\textbf{Abdulrahman Algunayah} holds a B.Sc. degree in Electrical Engineering from King Saud University. He is Infrastructure Design Engineer at Saudi Telecom (STC).

 \newpage
 
\startfigure\includegraphics[width=0.4\columnwidth]{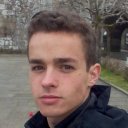} 
\end{figure}\textbf{Jorge Martín-Pérez} is a telematics engineering Ph.D. student at Universidad Carlos III de Madrid (UC3M). He obtained a B.Sc. in mathematics and computer science, both in 2016 from Universidad Autónoma de Madrid. In 2017, he obtained an M.Sc. in telematics engineering from UC3M. His research focuses on optimal resource allocation in 5G networks, and since 2016 he participates in EU-funded research projects.\ITUpar

\startfigure\includegraphics[width=0.4\columnwidth]{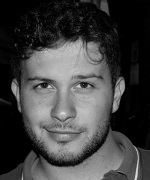} 
\end{figure}\textbf{Luigi Girletti} is a telematics engineering Ph.D. student at Universidad Carlos III de Madrid (UC3M). He obtained both a B.Sc. and an M.Sc in computer science engineering in 2014 and 2018 from Universitá degli Studi Federico II di Napoli. His research focuses on SDN/NFV, Artificial Intelligence applied to next-generation networks, and 5G-based EU-funded research projects.\ITUpar

\startfigure\includegraphics[width=0.4\columnwidth]{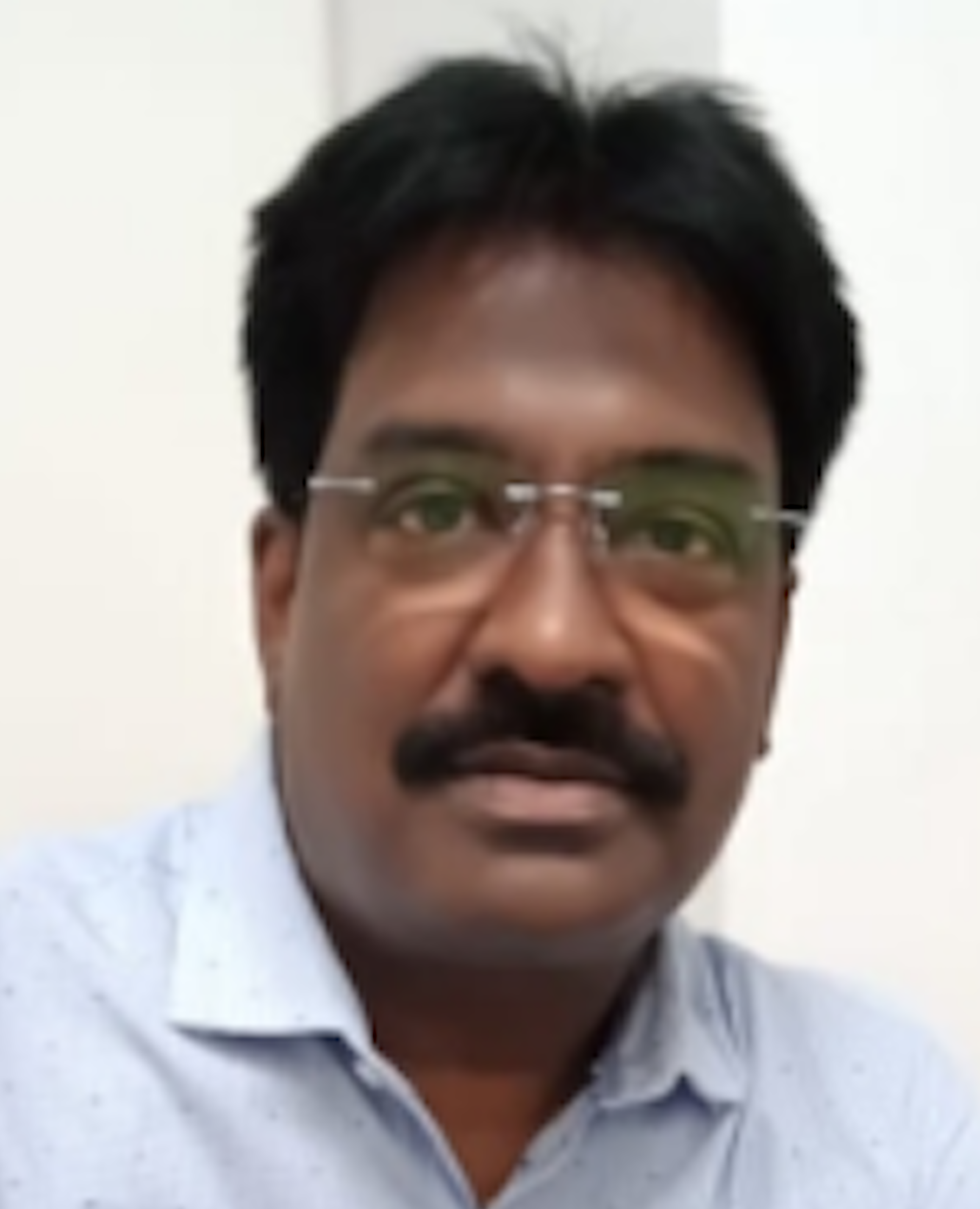} 
\end{figure}\textbf{Rajasekar Mohan}, alumnus of Indian Institute of Technology, Madras, is pursuing his research in the domain of wireless communication at PES University.  His research interests include embedded systems, IoT, Robotics and application of ML in communication.  He has served in the Indian Air Force for over 23 years in various technology roles in the field of communication.  Currently, he is an Associate Professor in the department of ECE at PES University, Bangalore.\ITUpar

\startfigure\includegraphics[width=0.4\columnwidth]{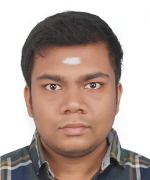} 
\end{figure}\textbf{K Venkat Ramnan} is an undergrad student in the Electronics and Communication Department at PES University, India. He is a beginner in Machine Learning, Deep Learning and Image processing. He also works with the Internet of things, Networking and Linux. He aims to apply Machine learning and Deep Learning in various fields including Wireless Communication, Healthcare and various other domains of Electronics and Communication.\ITUpar

\startfigure\includegraphics[width=0.4\columnwidth]{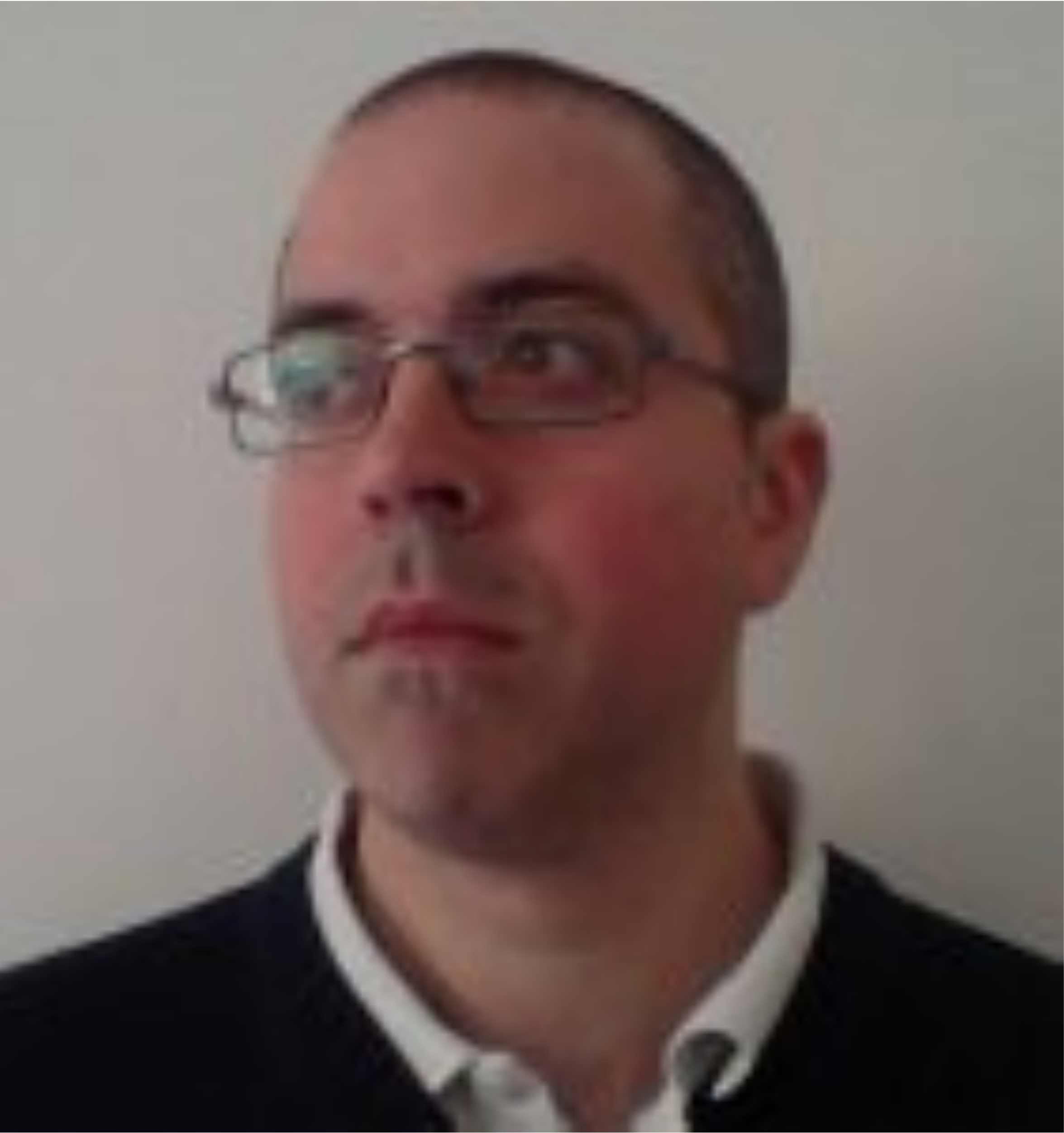} 
\end{figure}\textbf{Boris Bellalta} is an Associate Professor in the Department of Information and Communication Technologies (DTIC) at Universitat Pompeu Fabra (UPF). He is the head of the Wireless Networking research group at DTIC/UPF. His research interests are in the area of wireless networks, with emphasis on the design and performance evaluation of new architectures and protocols.
\ITUpar

\end{document}